\documentclass[11pt]{article}\include{ddefs}
\usepackage{amsmath}
\usepackage{graphics,pifont,array}
\usepackage{graphicx}
\usepackage{endnotes}
\usepackage{physics}
\begin{document}

\begin{titlepage}

\begin{center}

{\Large{        \mbox{   }                          \\
                \mbox{   }                          \\
                \mbox{   }                          \\
                \mbox{   }                          \\
                \mbox{   }                          \\
                \mbox{   }                          \\
                \mbox{   }                          \\
                \mbox{   }                          \\
                \mbox{   }                          \\
                \mbox{   }                          \\
        {\textbf{WIGNER-WILKINS NEUTRON/NUCLEUS SCATTERING  \\
    KERNEL QUANTUM-MECHANICALLY DERIVED } }          \\  		
               \mbox{    }                          \\
               \mbox{    }                          \\
              J. A. Grzesik                         \\
           Allwave Corporation                      \\
        3860 Del Amo Boulevard                      \\
                Suite 404                           \\
           Torrance, CA 90503                       \\
                \mbox{    }                         \\
           (818) 749-3602                           \\ 
            jan.grzesik@hotmail.com                 \\
              \mbox{     }                          \\
              \mbox{     }                          \\  
              \today                                      }  }

\end{center}

\end{titlepage}

\setcounter{page}{2}

\pagenumbering{roman}

\setcounter{page}{2}

\vspace*{+2.225in}

\begin{abstract}

	\parindent=0.245in

	We undertake herein to derive the Wigner-Wilkins [W-W] neutron/nucleus scattering kernel,
	a foundation stone in neutron thermalization theory, on the basis of a self-contained
	calculation in quantum mechanics.  Indeed, a quantum-mechanical derivation of the W-W kernel
	is available in the literature, cited below, but it is, in our opinion, robbed of
	conviction by being couched in terms of an excessive generality.  Here, by contrast, we
	proceed along a self-contained route relying on the Fermi pseudopotential and a
	first-order term in a time-dependent Born approximation series.  Our calculations are fully
	explicit at every step and, in particular, we tackle in its every detail a final
	integration whose result is merely stated in the available literature.  Furthermore,
	and perhaps the most important point of all, we demonstrate that the quantum-mechanical
	W-W kernel outcome is identical down to the last iota with its classical antecedent,
	classical not only by virtue of historical precedence but also by being based
	on classical Newtonian mechanics.

\end{abstract}

\pagestyle{plain}

\parindent=0.5in

\newpage

\pagenumbering{arabic}

\pagestyle{myheadings}

\setlength{\parindent}{0pt}

\pagestyle{plain}

\parindent=0.5in

\newpage
\mbox{   }

\pagestyle{myheadings}

\parindent=0.5in

\pagenumbering{arabic}

\pagestyle{myheadings}

\setcounter{page}{1}


\markright{J. A. Grzesik \\ wigner-wilkins scattering kernel quantum-mechanically derived}
\section{Introduction}
\vspace{-2mm}
    The pages that follow do not seek to break any new ground.  Rather, they seek to aggregate,
    at the very least by mere literature allusion,
    the available computations of the Wigner-Wilkins neutron-nucleus scattering kernel which
    occupies a legendary position in the theory of neutron thermalization [1].  Astoundingly
    enough, it turns out that, be the physical perspective classical or quantum mechanical,
    and despite an endless torrent of entirely dissimilar, confoundingly intricate analytic
    steps, the end results turn out to be identical, a most striking confluence indeed which
    motivates our ensuing discourse.  Of course, the classical/quantum agreement, while it
    is most welcome and not entirely unexpected, is nevertheless not {\emph{a priori}} guaranteed.
    \vspace{-0mm}    
    
    The foundational W-W document [1] proceeds from a purely classical basis to consider neutron
    scattering from nuclei distributed in their velocity magnitudes $v_{2}$ in accordance with the
    standard Maxwell-Boltzmann [M-B] law
    \begin{equation}
            {\cal{N}}\left(\frac{mA}{2\pi kT}\right)^{3/2}e^{-mAv_{2}^{2}/2kT}\,,
    \end{equation}
    with $m$ being the neutron mass, $mA$ that of each thermalizing nucleus ($A$ being the
    corresponding nuclear mass number), and both nuclear density ${\cal{N}}$ and temperature $T$
    regarded as spatially uniform.  Symbol $k$ is the Boltzmann constant.  The scattering
    kernel at issue, idiosyncratically denoted in [1] as ${\underbar{P}}(v,v_{1})$ (with $v_{1}$
    as the initial laboratory frame neutron speed, and $v$ as it post-collision outcome),
    complete with an underbar, is gotten by regarding the neutron-nucleus impact to yield
    an isotropically distributed pair in the two-particle center of mass frame.  The
    development in [1] is fully explicit save for the required integration over the cosine
    $\mu$ between incoming/outgoing neutron velocities $\mbox{\boldmath$v$}_{1}$ and $\mbox{\boldmath$v$}$
    respectively.  For these cumbersome integrations
    only a trend is suggested, a trend filled many years later in [2], [3], and, from the
    pen of the undersigned, in [4].  Reference [4] provides an alternative
    mapping of variables which greatly streamlines the requisite integrations and leads
    to a result having a high degree of symmetry between speeds $v_{1}$ and $v,$ fully equivalent,
    of course, with what is on view in [1]-[3].  Without further ado we recite at this point the
    dimensionless structure of
    ${\underbar{P}}(v,v_{1})$ (Eqs. (5) and (5a) in [1]):
 \begin{equation}
	{\underbar{P}}(v,v_{1}) = \frac{2\theta^{2}v}{\sqrt{\pi}v_{1}}\left\{\rule{0mm}{1.2cm}\begin{array}{l}
		 e^{m(v_{1}^{2}-v^{2})/2kT}
		\left[\rule{0mm}{4mm}I\left(\rule{0mm}{4mm}\frac{\theta v_{1}-\zeta v}{\sqrt{2kT/m}}\right)
		+I\left(\rule{0mm}{4mm}\frac{\theta v_{1}+\zeta v}{\sqrt{2kT/m}}\right)\right]\!+\!
		\left[\rule{0mm}{4mm}I\left(\rule{0mm}{4mm}\frac{\theta v-\zeta v_{1}}{\sqrt{2kT/m}}\right)-
		\rule{0mm}{4mm}I\left(\rule{0mm}{4mm}\frac{\theta v+\zeta v_{1}}{\sqrt{2kT/m}}\right)\right] \\    
		\\
		 e^{m(v_{1}^{2}-v^{2})/2kT}
		\left[\rule{0mm}{4mm}I\left(\rule{0mm}{4mm}\frac{\theta v_{1}-\zeta v}{\sqrt{2kT/m}}\right)
		-I\left(\rule{0mm}{4mm}\frac{\theta v_{1}+\zeta v}{\sqrt{2kT/m}}\right)\right]\!+\!
		\left[\rule{0mm}{4mm}I\left(\rule{0mm}{4mm}\frac{\theta v-\zeta v_{1}}{\sqrt{2kT/m}}\right)+
		\rule{0mm}{4mm}I\left(\rule{0mm}{4mm}\frac{\theta v+\zeta v_{1}}{\sqrt{2kT/m}}\right)\right] \\   
	\end{array} \right.                             
\end{equation}
respectively for $v_{1}<v$ and $v_{1}>v,$ and wherein
\begin{equation}
	\left\{\begin{array}{rcl}
		    \theta & = & (A+1)/2\sqrt{A} \\
		    \zeta  & = & (A-1)/2\sqrt{A}
		    \end{array} \right.       
\end{equation}
whereas function 
\begin{equation}
  I(\xi) = \int_{0}^{\,\xi}e^{-t^{2}}dt\,,
\end{equation}
odd in respect to its argument $\xi,$ is the unnormalized error function, a direct inheritance from its M-B parent in (1).
\newpage
\mbox{   }
\newline

    Now, while it is true that the literature does offer a quantum-mechanical derivation of (2) [5], that particular derivation
    seems to be mired, on the one hand, by a context of excessive generality allowing the forest to be seen at the expense of
    its trees, while, on the other, it gives once again only the final outcome of the requisite integration over the cosine
    $\mu$ of the angle separating the incoming/outgoing neutron flight directions.  That particular integration, differing entirely
    from its double counterpart at the base of (2), is remarkable, at least for the present author, by finding its antiderivative
    amid an interplay of suitably tailored variable clusters, already seen in (2), themselves situated as upper limits of the
    integral (4) defining the unnormalized error
    function.  By contrast, we intend to pursue a somewhat more humble, more modest path explicitly utilizing the Fermi
    pseudopotential [6]-[8] and resting content with just the first Born scattering approximation.\footnote{In all fairness,
    this direct contact, Dirac delta pseudopotential {\emph{is}} mentioned toward the end of [5], on its pp. 534-535.}  Moreover, we fully
    intend to plough through that final $\mu$ integration, both by reason of its pedagogical value and its {\emph{sine qua non}}
    burden of credibility.
    \vspace{-3mm}
    \section{Neutron/nucleus scattering in a Fermi pseudopotential/Born approximation framework}
    \vspace{-3mm}
        Henceforth we utilize subscripts $n$ and $\nu$ respectively for neutron and nucleus, and similarly $i$ and $f$ for
         initial/final states.\footnote{One needs no reminding that $i$ as a suffix has nothing whatsoever to do with the
         imaginary unit elsewhere utilized in its normal capacity.}  As needed, these subscripts will be staggered, and then,
         at the end, once the integration over
         the M-B nuclear thermalizing background has been performed, neutron suffix $n$ will be dropped by reason of
         having become superfluous.  We adopt a strictly contact interaction between neutron and nucleus, encapsulated
         in a Fermi pseudopotential [6]-[8]
         \begin{equation}
         	U({\bf{r}}_{n},{\bf{r}}_{\nu})=\frac{2\pi\hbar^{2}a_{fr}(A+1)}{mA}\delta({\bf{r}}_{n}-{\bf{r}}_{\nu})\,,
         \end{equation}      
         with $a_{fr}$ being the so-called {\emph{``free''}} scattering length,\footnote{Scattering length
         $a_{fr}$ is negative when there
         exists no bound neutron/nucleus state [6], and positive otherwise.  This physical option is without bearing on our final goal,
         since length $a_{fr}$ gauges the {\emph{``zero energy''}} scattering cross-section $\sigma_{0}$ in accordance with
         	$\sigma_{0}=4\pi a_{fr}^{2},$ altogether indifferent to the sign of $a_{fr}.$  One may note in passing that [8]
         hews to the opposite sign convention as regards $a_{fr},$ and indeed replaces it with the symbol $b_{fr},$ positive
         when the neutron/nucleus interaction is repulsive and thus inherently incapable of bound-state entrapment, all of which is designed
         to sow confusion in an already somewhat murky corner of nuclear theory.}  $\delta$ the three-dimensional Dirac delta, and factor
         $(A+1)/A$ conveying the reduced neutron/nucleus mass accompanying passage from laboratory to center-of-mass frames ({\emph{cf.}}
         [6], especially pp. 73-77).
         
             As the incoming, time-dependent, unperturbed wave function $\psi_{i}(\mbox{\boldmath$r$}_{n},\mbox{\boldmath$r$}_{\nu},t)$
             we take
             \begin{equation}
             \psi_{i}(\mbox{\boldmath$r$}_{n},\mbox{\boldmath$r$}_{\nu},t) = \frac{1}{V}\exp\left\{\frac{i}{\hbar}\left(
               \mbox{\boldmath$p$}_{ni}\mbox{\boldmath$\cdot\, r$}_{n}+\mbox{\boldmath$p$}_{\nu i}\mbox{\boldmath$\cdot\, r$}_{\nu}
        -\frac{1}{2m}\!\left[\rule{0mm}{4mm}\,\mbox{\boldmath$p$}_{ni}^{2}+\mbox{\boldmath$p$}_{\nu i}^{2}/A\,\right]t\right)\right\}\,,
             \end{equation}
         a product of individual, free-particle energy/momentum eigenfunctions, each normalized to unity within some region of sufficiently
         large volume $V.$  Symbol $\mbox{\boldmath$p$}$ stands as always for vector momentum whereas its square, here and below, is a
         shorthand for the inner product
         \begin{equation}
         \mbox{\boldmath$p$}^{2}=\mbox{\boldmath$p\,\cdot\, p$} 
         \end{equation}
         \newpage
         \mbox{    }
         \newline
         \newline
         \newline
         with appropriate subscripts appended.  The space-time evolution of (6) is governed by the unperturbed Hamiltonian
         \begin{equation}
         H_{0} = -\frac{\hbar^{2}}{2m}{\mbox{\boldmath$\nabla$}_{n}}^{\!\!\!2}-\frac{\hbar^{2}}{2mA}{\mbox{\boldmath$\nabla$}_{\nu}}^{\!\!\!2}
         \end{equation}
         and, with its use, one can present the scattered, first-order Born, additive perturbation $\psi_{s}$ of $\psi_{i},$ induced by
         the neutron/nucleus impacts implied by $U({\bf{r}}_{n},{\bf{r}}_{\nu}),$ as
         \begin{equation}
          \psi_{s}(\mbox{\boldmath$r$}_{n},\mbox{\boldmath$r$}_{\nu},t) = \frac{1}{i\hbar}e^{-\frac{i}{\hbar}H_{0}t}\int_{-\infty}^{\,t}
           e^{\frac{i}{\hbar}H_{0}\tau}U({\bf{r}}_{n},{\bf{r}}_{\nu}) \psi_{i}(\mbox{\boldmath$r$}_{n},\mbox{\boldmath$r$}_{\nu},\tau)\,d\tau\,.                 
         \end{equation}
         An integrated term on the left at $\tau\rightarrow -\infty$ has naturally been dropped, and we further assume the presence
         of some sort of a {\emph{deus ex machina}} physical attenuation so as to assure convergence on the right in that same limit
         $\tau\rightarrow -\infty.$
         
         We allow next an essentially infinite time $t\rightarrow\infty$ to elapse for a {\emph{``sufficiently mature''}} neutron/nucleus
         interaction to be fully consummated, and project scattered state (6) onto
         \begin{equation}
       	\psi_{f}(\mbox{\boldmath$r$}_{n},\mbox{\boldmath$r$}_{\nu},t) = \frac{1}{(2\pi\hbar)^{3}}\exp\left\{\frac{i}{\hbar}\left(
\mbox{\boldmath$p$}_{nf}\mbox{\boldmath$\cdot\, r$}_{n}+\mbox{\boldmath$p$}_{\nu f}\mbox{\boldmath$\cdot\, r$}_{\nu}\right)\right\}\,,
         \end{equation}
     an eigenstate of the final, outgoing neutron/nucleus momenta $\mbox{\boldmath$p$}_{nf}$ and $\mbox{\boldmath$p$}_{\nu f},$
     normalized in accordance with the viewpoint that both these latter admit a continuum of values.\footnote{The projection of two such states upon each other is a product of two Dirac deltas having as arguments the differences of their respective neutron/nucleaus momenta.}  And, when forming such a projection, $\bra{\psi_{f}}\ket{\psi_{s}},$ we allow both Hamiltonians in (9), by virtue of
     their being Hermitian, to operate sinistrally upon $\psi_{f}.$  After $\mbox{\boldmath$r$}_{n}$ and $\mbox{\boldmath$r$}_{\nu}$
     have been duly clamped into strict unison by virtue of (5), and once all remaining space-time ($\mbox{\boldmath$r$}_{n},t)$ integrations have thus metamorphosed into the appropriate Dirac deltas, one finds
     \begin{eqnarray}
     \bra{\psi_{f}}\ket{\psi_{s}} & \approx\atop{t\rightarrow\infty}& -i\,\frac{8\pi^{2}\hbar^{2}a_{fr}(A+1)}{VA}     
     \exp\left\{\frac{i}{2\hbar m}
     \left(\rule{0mm}{4mm}\,\mbox{\boldmath$p$}_{nf}^{2}+\mbox{\boldmath$p$}_{\nu f}^{2}/A\,\right)t\right\}\times \nonumber \\
          &  & \rule{1.5cm}{0mm} \delta\!\left(\rule{0mm}{4mm}\mbox{\boldmath$p$}_{nf}+\mbox{\boldmath$p$}_{\nu f}
          -\mbox{\boldmath$p$}_{ni}-\mbox{\boldmath$p$}_{\nu i}\right)
            \delta\!\left(\rule{0mm}{4mm}\mbox{\boldmath$p$}_{nf}^{2}+\mbox{\boldmath$p$}_{\nu f}^{2}/A
          -\mbox{\boldmath$p$}_{ni}^{2}-\mbox{\boldmath$p$}_{\nu i}^{2}/A\right)\,,   
     \end{eqnarray}
 wherein the two delta functions conserve the composite neutron/nucleus vector momentum and the composite energy, in that order.
 
       Since our next objective is not $ \bra{\psi_{f}}\ket{\psi_{s}}$ {\emph{per se}} but rather its absolute square
       $| \bra{\psi_{f}}\ket{\psi_{s}}|^{2},$ we are immediately forced to confront bewildering constructs such as
       \begin{equation}
       	\delta\!\left(\rule{0mm}{4mm}\mbox{\boldmath$p$}_{nf}+\mbox{\boldmath$p$}_{\nu f}
       	       - \mbox{\boldmath$p$}_{ni}-\mbox{\boldmath$p$}_{\nu i}\right)^{2}
       \end{equation}	    
with a view to extracting from them some physically meaningful information.  We do so by writing this seemingly ambiguous structure in
it hybrid form as
\begin{eqnarray}
	\lefteqn{
\delta\!\left(\rule{0mm}{4mm}\mbox{\boldmath$p$}_{nf}+\mbox{\boldmath$p$}_{\nu f}
    - \mbox{\boldmath$p$}_{ni}-\mbox{\boldmath$p$}_{\nu i}\right)^{2}  =  
  \delta\!\left(\rule{0mm}{4mm}\mbox{\boldmath$p$}_{nf}+\mbox{\boldmath$p$}_{\nu f}
     - \mbox{\boldmath$p$}_{ni}-\mbox{\boldmath$p$}_{\nu i}\right)\times }  \nonumber  \\
      &     & \rule{1.5cm}{0mm} \left(\frac{1}{2\pi\hbar}\right)^{3}
         \int_{all\;\mbox{\boldmath$r$}}\exp\left\{\frac{i}{\hbar}\left(\rule{0mm}{4mm}\mbox{\boldmath$p$}_{nf}+\mbox{\boldmath$p$}_{\nu f}
         - \mbox{\boldmath$p$}_{ni}-\mbox{\boldmath$p$}_{\nu i}\right)\mbox{\boldmath$\cdot\,r$}\right\}d\mbox{\boldmath$r$}\,,		
\end{eqnarray}
\newpage
\mbox{   }
\newline
\newline
\newline
the differential $d\mbox{\boldmath$r$}$ being a standard shorthand for the product $dx\,dy\,dz.$  And then, on the strength of a delta
function being already present as a factor on the right, we can simply replace the exponential by unity and
thus arrive at\footnote{Manipulations of this sort, in the words of the acclaimed physicist Steven Weinberg, are sure to bring tears
	to mathematicians' eyes.  Comments designed to provide some credibility to maneuvers of this
sort, and to assuage the intellectual anxiety which they provoke, can be found in [9]-[10].}
\begin{equation}
\delta\!\left(\rule{0mm}{4mm}\mbox{\boldmath$p$}_{nf}+\mbox{\boldmath$p$}_{\nu f}
   - \mbox{\boldmath$p$}_{ni}-\mbox{\boldmath$p$}_{\nu i}\right)^{2}  =  V\left(\frac{1}{2\pi\hbar}\right)^{3}
        \delta\!\left(\rule{0mm}{4mm}\mbox{\boldmath$p$}_{nf}+\mbox{\boldmath$p$}_{\nu f}
              - \mbox{\boldmath$p$}_{ni}-\mbox{\boldmath$p$}_{\nu i}\right)\,.
\end{equation}
and similarly  
\begin{eqnarray}
	\lefteqn{\delta\!\left(\rule{0mm}{4mm}\mbox{\boldmath$p$}_{nf}^{2}+\mbox{\boldmath$p$}_{\nu f}^{2}/A
		 -\mbox{\boldmath$p$}_{ni}^{2}-\mbox{\boldmath$p$}_{\nu i}^{2}/A\right)^{2} =   
	\delta\!\left(\rule{0mm}{4mm}\mbox{\boldmath$p$}_{nf}^{2}+\mbox{\boldmath$p$}_{\nu f}^{2}/A
	 -\mbox{\boldmath$p$}_{ni}^{2}-\mbox{\boldmath$p$}_{\nu i}^{2}/A\right) \times }  \nonumber   \\
 	&   &  \rule{5.3cm}{0mm} \left(\frac{1}{4\pi m\hbar}\right)
 		\int_{-\infty}^{\,\infty}\exp\left\{\frac{i}{2m\hbar}\left(\rule{0mm}{4mm}\mbox{\boldmath$p$}_{nf}^{2}+\mbox{\boldmath$p$}_{\nu f}^{2}/A
 		- \mbox{\boldmath$p$}_{ni}^{2}-\mbox{\boldmath$p$}_{\nu i}^{2}/A\right)\tau\right\}d\tau  \nonumber  \\
 	&  & \rule{4.5cm}{0mm}= D\left(\frac{1}{4\pi m\hbar}\right)\delta\!\left(\rule{0mm}{4mm}\mbox{\boldmath$p$}_{nf}^{2}+\mbox{\boldmath$p$}_{\nu f}^{2}/A
 		 -\mbox{\boldmath$p$}_{ni}^{2}-\mbox{\boldmath$p$}_{\nu i}^{2}/A\right)\,,  
\end{eqnarray}    
	with $D$ being some sufficiently large interaction epoch ($D$ for {\emph{``duration''}}).  Altogether then we get from
	(11), (14), and (15) combined that\footnote{Taking full advantage once more of the relaxed, indulgent attitude which physicists arrogate
		to themselves {\emph{vis-\`{a}-vis}} mathematics, we have simply promoted to full equality the approximate status in (11), itself
		resting on mathematically shaky grounds.} 
	\begin{eqnarray}
		\left|\rule{0mm}{3.5mm}\bra{\psi_{f}}\ket{\psi_{s}}\right|^{2} & = & \frac{2a_{fr}^{2}(A+1)^{2}D}{mVA^{2}}
		 \delta\!\left(\rule{0mm}{4mm}\mbox{\boldmath$p$}_{nf}+\mbox{\boldmath$p$}_{\nu f}	
		  - \mbox{\boldmath$p$}_{ni}-\mbox{\boldmath$p$}_{\nu i}\right)
		  	 \delta\!\left(\rule{0mm}{4mm}\mbox{\boldmath$p$}_{nf}^{2}+\mbox{\boldmath$p$}_{\nu f}^{2}/A
		  	 -\mbox{\boldmath$p$}_{ni}^{2}-\mbox{\boldmath$p$}_{\nu i}^{2}/A\right)\,.
	\end{eqnarray}

     With (16) in hand, the transition rate 
     $R(\mbox{\boldmath$p$}_{ni}\rightarrow \mbox{\boldmath$p$}_{nf}|\,\mbox{\boldmath$p$}_{\nu i}\rightarrow \mbox{\boldmath$p$}_{\nu f})$
     per combined outgoing momentum interval $d\mbox{\boldmath$p$}_{nf}\times\, d\mbox{\boldmath$p$}_{\nu f}$ around
     $(\mbox{\boldmath$p$}_{nf},\mbox{\boldmath$p$}_{\nu f})$ reads
     \begin{eqnarray}
     R(\mbox{\boldmath$p$}_{ni}\rightarrow \mbox{\boldmath$p$}_{nf}|\,\mbox{\boldmath$p$}_{\nu i}\rightarrow \mbox{\boldmath$p$}_{\nu f}) & = &
           \left|\rule{0mm}{3.5mm}\bra{\psi_{f}}\ket{\psi_{s}}\right|^{2}/\,D\,,         	
     \end{eqnarray}
    whereupon one further division by the incoming neutron flux $p_{ni}/mV$ yields a differential transition cross section
    $d\sigma(\mbox{\boldmath$p$}_{ni}\rightarrow \mbox{\boldmath$p$}_{nf}|\,\mbox{\boldmath$p$}_{\nu i}\rightarrow \mbox{\boldmath$p$}_{\nu f})$
    in the form\footnote{Under most normal circumstances one would be obliged to divide by the {\emph{relative}} velocity
    	$|\mbox{\boldmath$p$}_{ni}-\mbox{\boldmath$p$}_{\nu i}/A|/m$ prior to impact, and not merely, as here, $p_{ni}/m.$  The physical basis for such
    	analytic luxury is the contact nature of the pseudoptential interaction.  One may note, however, that uniformly across all four
    	references [1]-[4], a {\emph{bona fide}} relative velocity is duly utilized, only to be conveniently absorbed by the remainder
    	of the analytic apparatus there encountered.}   
    \begin{eqnarray}
    d\sigma(\mbox{\boldmath$p$}_{ni}\rightarrow \mbox{\boldmath$p$}_{nf}|\,\mbox{\boldmath$p$}_{\nu i}\rightarrow \mbox{\boldmath$p$}_{\nu f}) & = &
      \frac{2a_{fr}^{2}(A+1)^{2}}{p_{ni}A^{2}}
      \delta\!\left(\rule{0mm}{4mm}\mbox{\boldmath$p$}_{nf}+\mbox{\boldmath$p$}_{\nu f}   	
         - \mbox{\boldmath$p$}_{ni}-\mbox{\boldmath$p$}_{\nu i}\right)\times    \\
       &  & \rule{2.2cm}{0mm} \delta\!\left(\rule{0mm}{4mm}\mbox{\boldmath$p$}_{nf}^{2}+\mbox{\boldmath$p$}_{\nu f}^{2}/A
                 -\mbox{\boldmath$p$}_{ni}^{2}-\mbox{\boldmath$p$}_{\nu i}^{2}/A\right)
                      p_{nf}^{2}\,dp_{nf}\,d\Omega_{\mbox{\boldmath$p$}_{nf}} d\mbox{\boldmath$p$}_{\nu f}\,,  \nonumber
    \end{eqnarray}
with $d\Omega_{\mbox{\boldmath$p$}_{nf}}$ being an increment of solid angle around $\mbox{\boldmath$p$}_{nf}.$  Of course, we have no
interest whatsoever in $\mbox{\boldmath$p$}_{\nu f},$ indifference expressed by simply integrating over
its full range, with the result
\newpage
\mbox{   }
\newline
 \begin{eqnarray}
	\frac{d^{\,2}\sigma(\mbox{\boldmath$p$}_{ni}\rightarrow \mbox{\boldmath$p$}_{nf}|\,\mbox{\boldmath$p$}_{\nu i})}
	   {dp_{nf}\,d\Omega_{\mbox{\boldmath$p$}_{nf}}}
	 & = &
	\frac{2a_{fr}^{2}p_{nf}^{2}(A+1)^{2}}{p_{ni}A^{2}}\,	
	        \delta\!\left(\rule{0mm}{4mm}\mbox{\boldmath$p$}_{nf}^{2}+
	\left\{\mbox{\boldmath$p$}_{ni}+\mbox{\boldmath$p$}_{\nu i}-\mbox{\boldmath$p$}_{nf}\right\}^{2}\!\!/A
	-\mbox{\boldmath$p$}_{ni}^{2}-\mbox{\boldmath$p$}_{\nu i}^{2}/A\right)  
\end{eqnarray}
or else, more explcitly
\begin{eqnarray}
	\lefteqn{
	\frac{d^{\,2}\sigma(\mbox{\boldmath$p$}_{ni}\rightarrow \mbox{\boldmath$p$}_{nf}|\,\mbox{\boldmath$p$}_{\nu i})}
	{dp_{nf}\,d\Omega_{\mbox{\boldmath$p$}_{nf}}} = \frac{2a_{fr}^{2}p_{nf}^{2}(A+1)^{2}}{p_{ni}A^{2}}\,\times }    \\
	&   & \rule{2.9cm}{0mm}	
	\delta\!\left(\rule{0mm}{7mm}\left(\frac{A+1}{A}\right)\mbox{\boldmath$p$}_{nf}^{2}-
	\left(\frac{A-1}{A}\right)\mbox{\boldmath$p$}_{ni}^{2}-     
	\left(\frac{2}{A}\right)\mbox{\boldmath$p$}_{nf}\mbox{\boldmath$\,\cdot\,\,$}\mbox{\boldmath$p$}_{ni}-
\left(\frac{2}{A}\right)\mbox{\boldmath$p$}_{\nu i}\mbox{\boldmath$\,\cdot$}\left\{\rule{0mm}{4mm}\mbox{\boldmath$p$}_{nf}-\mbox{\boldmath$p$}_{ni}\right\}\right)\,.\nonumber
\end{eqnarray}

     Our next task will be to average (20) over the thermalizing nuclear Maxwellian
     \begin{equation}
     	\left(\frac{1}{2\pi mAkT}\right)^{3/2}\exp\left(-\frac{p_{\nu i}^{2}}{2mAkT}\right)\,,
     \end{equation}
 and, with this in mind, it becomes essential that we extricate $\mbox{\boldmath$p$}_{\nu i}$ from within the delta argument out
 into the open so as to render it susceptible to the intended integration weighted by (21).  This is easily done by invoking once
 again the standard connection between Dirac's delta and a Fourier integral.  Thus
 \begin{eqnarray}
 	\lefteqn{
 		\frac{d^{\,2}\sigma(\mbox{\boldmath$p$}_{ni}\rightarrow \mbox{\boldmath$p$}_{nf}|\,\mbox{\boldmath$p$}_{\nu i})}
 		{dp_{nf}\,d\Omega_{\mbox{\boldmath$p$}_{nf}}} = \frac{a_{fr}^{2}p_{nf}^{2}(A+1)^{2}}{\pi p_{ni}A^{2}}\,\times }     \\
 	&   & \rule{0.3cm}{0mm}\int_{-\infty}^{\,\infty}\exp\left\{\rule{0mm}{8mm}i	
 	\left(\rule{0mm}{7mm}\left(\frac{A+1}{A}\right)\mbox{\boldmath$p$}_{nf}^{2}-
 	\left(\frac{A-1}{A}\right)\mbox{\boldmath$p$}_{ni}^{2}-     
 	\left(\frac{2}{A}\right)\mbox{\boldmath$p$}_{nf}\mbox{\boldmath$\,\cdot\,\,$}\mbox{\boldmath$p$}_{ni}-
 	\left(\frac{2}{A}\right)\mbox{\boldmath$p$}_{\nu i}\mbox{\boldmath$\,\cdot$}
 	\left\{\rule{0mm}{4mm}\mbox{\boldmath$p$}_{nf}-\mbox{\boldmath$p$}_{ni}\right\}\right)\tau\right\}d\tau \,, \nonumber  
 \end{eqnarray}
variable $\tau$ having in this instance no particular physical meaning, save for the requirement of being equipped with the units
of $time^{2}/(mass \times distance)^{2}$ in order to maintain in (22) a dimensional balance, left and right.
\section{The M-B thermalizing average}
     M-B thermalization at temperature $T$ in accordance with (21) forces us to confront next
\begin{eqnarray}
	\lefteqn{\left\langle
		\frac{d^{\,2}\sigma(\mbox{\boldmath$p$}_{ni}\rightarrow \mbox{\boldmath$p$}_{nf})}
		{dp_{nf}\,d\Omega_{\mbox{\boldmath$p$}_{nf}}}\right\rangle_{\!\!T} = \left(\frac{a_{fr}^{2}p_{nf}^{2}(A+1)^{2}}{\pi p_{ni}A^{2}}\right)
		                                        \left(\rule{0mm}{7mm}\frac{1}{2\pi mAkT}\right)^{3/2}\!\times }     \\
	&   & \rule{2.2cm}{0mm}\int_{-\infty}^{\,\infty}d\tau\exp\left\{\rule{0mm}{8mm}i	
	\left(\rule{0mm}{7mm}\left(\frac{A+1}{A}\right)\mbox{\boldmath$p$}_{nf}^{2}-
	\left(\frac{A-1}{A}\right)\mbox{\boldmath$p$}_{ni}^{2}-     
	\left(\frac{2}{A}\right)\mbox{\boldmath$p$}_{nf}\mbox{\boldmath$\,\cdot\,\,$}\mbox{\boldmath$p$}_{ni}\right)\tau\right\}\,\times \nonumber  \\
	&   & \rule{4.8cm}{0mm} \int_{all \; \mbox{\boldmath$p$}_{\nu i}}\exp\left\{\rule{0mm}{8mm}-\left(\rule{0mm}{7mm}\frac{p_{\nu i}^{2}}{2mAkT}+
	\left(\frac{2i\tau}{A}\right)\mbox{\boldmath$p$}_{\nu i}\mbox{\boldmath$\,\cdot$}
	\left\{\rule{0mm}{4mm}\mbox{\boldmath$p$}_{nf}-\mbox{\boldmath$p$}_{ni}\right\}\right)\right\}\,d\mbox{\boldmath$p$}_{\nu i} \,. \nonumber  
\end{eqnarray}
\newpage
\mbox{  }
\newline
\newline
\newline
Index $\nu$ being no longer in need of any in/out qualification, we drop its $i$ and observe that the composite integral over all $\mbox{\boldmath$p$}_{\nu}$
splits into a product of three similar structures, each of which readily succumbs to square completion in its exponent and a subsequent
quadrature as a standard Gaussian once provision is made for a suitable path displacement in a fixed imaginary amount.
So\footnote{Alternatively, since
\begin{eqnarray}
	\int_{4\pi}\exp\left\{\rule{0mm}{7mm}-i\left(\frac{2\tau}{A}\right)\mbox{\boldmath$p$}_{\nu}\mbox{\boldmath$\,\cdot$}
	\left(\rule{0mm}{4mm}\mbox{\boldmath$p$}_{nf}-\mbox{\boldmath$p$}_{ni}\right)\right\}
	d\Omega_{\mbox{\boldmath$p$}_{\nu}} & = &
	 \frac{2\pi A\sin\left(\rule{0mm}{3mm}2\tau p_{\nu}|\mbox{\boldmath$p$}_{nf}-\mbox{\boldmath$p$}_{ni}|/A\right)}
	{\tau p_{\nu}|\mbox{\boldmath$p$}_{nf}-\mbox{\boldmath$p$}_{ni}|}\,, \nonumber
\end{eqnarray}
it follows that
\begin{eqnarray}
	\lefteqn{
		\int_{all \; \mbox{\boldmath$p$}_{\nu}}\exp\left\{\rule{0mm}{8mm}-\left(\rule{0mm}{7mm}\frac{p_{\nu}^{2}}{2mAkT}+
		\left(\frac{2i\tau}{A}\right)\mbox{\boldmath$p$}_{\nu}\mbox{\boldmath$\,\cdot$}
		\left\{\rule{0mm}{4mm}\mbox{\boldmath$p$}_{nf}-\mbox{\boldmath$p$}_{ni}\right\}\right)\right\}\,d\mbox{\boldmath$p$}_{\nu}  =  }  \nonumber  \\
 & & \rule{1.5cm}{0mm}\left(\frac{\pi A } 
                {\tau |\mbox{\boldmath$p$}_{nf}-\mbox{\boldmath$p$}_{ni}|}\right)
\int_{-\infty}^{\,\infty}\exp\left(\rule{0mm}{5mm}-\frac{p_{\nu}^{2}}{2mAkT}\right)
    \sin\left(\rule{0mm}{4mm}2\tau p_{\nu}|\mbox{\boldmath$p$}_{nf}-\mbox{\boldmath$p$}_{ni}|/A\right)p_{\nu}dp_{\nu}  \nonumber  \\
 &  & \rule{4.5cm}{0mm}=2\pi mAkT   
        \int_{-\infty}^{\,\infty}\exp\left(\rule{0mm}{5mm}-\frac{p_{\nu}^{2}}{2mAkT}\right)
        \cos\left(\rule{0mm}{4mm}2\tau p_{\nu}|\mbox{\boldmath$p$}_{nf}-\mbox{\boldmath$p$}_{ni}|/A\right)dp_{\nu}  \nonumber     
\end{eqnarray}
and the rest proceeds pretty much like that in connection with (24) and, naturally, with the same final result.}
\begin{eqnarray}
	\lefteqn{
	\int_{all \; \mbox{\boldmath$p$}_{\nu}}\exp\left\{\rule{0mm}{8mm}-\left(\rule{0mm}{7mm}\frac{p_{\nu}^{2}}{2mAkT}+
	\left(\frac{2i\tau}{A}\right)\mbox{\boldmath$p$}_{\nu}\mbox{\boldmath$\,\cdot$}
	\left\{\rule{0mm}{4mm}\mbox{\boldmath$p$}_{nf}-\mbox{\boldmath$p$}_{ni}\right\}\right)\right\}\,d\mbox{\boldmath$p$}_{\nu}  =  }  \nonumber  \\
 & &\rule{5.3cm}{0mm}\left(\rule{0mm}{4mm}2\pi mAkT\right)^{3/2}
 \exp\left(\rule{0mm}{5mm}-\frac{2\tau^{2}mkT(\mbox{\boldmath$p$}_{nf}-\mbox{\boldmath$p$}_{ni})^{2}}{A}\right)\,.
\end{eqnarray}
 
 The time is now more than ripe to abandon the quantum-mechanical viewpoint and to clamber onto our quotidian, classical plateau.  Indeed, Planck's
 constant $\hbar$ has now completely evaporated.  Furthermore, since there remains no visible vestige of nuclear attributes, no useful purpose
 is served by having subscript $n$ call attention to neutron properties.  And so we discard it.  Moving on, we first displace momenta in favor of
 normalized energies, $p\rightarrow p^{2}/2mkT=E,$ and, with this in mind, pass also to a dimensionless $\tau'=2mkT\tau.$  Since also
 $dp_{f}=\sqrt{mkT/2E_{f}}\times dE_{f},$ we altogether get
 \begin{eqnarray}
 	\lefteqn{\left\langle
 		\frac{d^{\,2}\sigma(\mbox{\boldmath$p$}_{i}\rightarrow \mbox{\boldmath$p$}_{f})}
 		{dE_{f}\,d\Omega_{\mbox{\boldmath$p$}_{f}}}\right\rangle_{\!\!T} = \left(\frac{a_{fr}^{2}(A+1)^{2}}{2\pi A^{2}}\right)\sqrt{\frac{E_{f}}{E_{i}}}\,\times } \nonumber \\
 	&   & \rule{1.5cm}{0mm}\int_{-\infty}^{\,\infty}d\tau'\exp\left\{\rule{0mm}{8mm}i	
 	\left(\rule{0mm}{7mm}\left(\frac{A+1}{A}\right)E_{f}-\left(\frac{A-1}{A}\right)E_{i}-
 	\left(\frac{2}{A}\right)\mu \sqrt{E_{f}E_{i}}\right)\tau'\right\}\,\times        \\
  &  & \rule{4.5cm}{0mm}\exp\left\{\rule{0mm}{8mm}-\left(\rule{0mm}{7mm}\left(\frac{1}{A}\right)E_{f}+\left(\frac{1}{A}\right)E_{i}-
 	\left(\frac{2}{A}\right)\mu \sqrt{E_{f}E_{i}}\right)\tau'^{\,2}\right\}\,,  \nonumber 
 \end{eqnarray}
with $\mu=\mbox{\boldmath$p$}_{f}\mbox{\boldmath$\,\cdot\,\,$}\mbox{\boldmath$p$}_{i}/p_{f}p_{i}$ being the cosine of the angle
separating initial/final momenta.  There looms thus yet another Gaussian and yet another need for square completion in its exponent.
Toward this goal it is
\newpage
\mbox{   }
\newline
\newline
\newline
advantageous to set\footnote{Dimensionless energy difference $\beta$ should not be confused with the use of this symbol in [1]
	to denote $1/\sqrt{2kT}.$}
\begin{equation}
	    \alpha    =   \frac{E_{f}+E_{i}-2\mu\sqrt{E_{f}E_{i}}}{A} \geq 0 
\end{equation}
and
\begin{equation}	    
   \beta  =   E_{f}-E_{i}\,,
\end{equation}
and thus to encounter
\begin{eqnarray}
	\left\langle
		\frac{d^{\,2}\sigma(\mbox{\boldmath$p$}_{i}\rightarrow \mbox{\boldmath$p$}_{f})}
		{dE_{f}\,d\Omega_{\mbox{\boldmath$p$}_{f}}}\right\rangle_{\!\!T} & = & \left(\frac{a_{fr}^{2}(A+1)^{2}}{2\pi A^{2}}\right)\sqrt{\frac{E_{f}}{E_{i}}}\,
	            \int_{-\infty}^{\,\infty}d\tau'\exp\left\{\rule{0mm}{5mm}-\alpha\tau'^{\,2}+i(\alpha+\beta)\tau'\right\}d\tau'  \nonumber \\
	  & = & \left(\frac{a_{fr}^{2}(A+1)^{2}}{2\pi A^{2}}\right)\sqrt{\frac{\pi E_{f}}{E_{i}}}\,\exp(-\frac{\beta}{2})\times
	  \frac{\exp\left\{-\left(\rule{0mm}{3mm}\alpha+\beta^{2}/\alpha\right)\!/4\right\}}{\sqrt{\alpha}} \,,           
\end{eqnarray}
whence there follows the penultimate reduction
\begin{eqnarray}
	\left\langle
	\frac{d\sigma(\mbox{\boldmath$p$}_{i}\rightarrow \mbox{\boldmath$p$}_{f})}
	{dE_{f}}\right\rangle_{\!\!T} & = & \frac{\sigma_{0}}{4\pi}\left(\frac{A+1}{A}\right)^{2}\sqrt{\frac{\pi E_{f}}{E_{i}}}\,
	\exp(-\frac{\beta}{2})
	\int_{-1}^{\,1}\frac{\exp\left\{-\left(\rule{0mm}{3mm}\alpha+\beta^{2}/\alpha\right)\!/4\right\}}{\sqrt{\alpha}}\,d\mu\,.            
\end{eqnarray}
under the integration $2\pi\int_{-1}^{\,1}d\mu\ldots$ over all $4\pi$ steradians of $\Omega_{\mbox{\boldmath$p$}_{f}},$ and
a replacement of $a_{fr}^{2}$ by $\sigma_{0}/4\pi.$
\section{Final reduction into error function form}
   Although one must relinquish any hope of integrating (29) in closed form, one can arrive at a consolation
   prize of sorts wherein close-to-perfect antiderivatives are placed into the upper limits of error function
   (4).  Thus
   \begin{eqnarray}
   	\frac{d}{d\mu}I\left(\frac{\beta\pm\alpha}{2\sqrt{\alpha}}\right) & = & -\left(\frac{\sqrt{E_{f}E_{i}}}{2A}\right)\times
   		\left(-\frac{\beta}{\alpha}\pm 1\right)\times\exp\left(\mp\frac{\beta}{2}\right)
   	\times\frac{\exp\left\{-(\alpha+\beta^{2}/\alpha)/4\right\}}{\sqrt{\alpha}}
   \end{eqnarray}
and so
\begin{eqnarray}
\left(\frac{A}{\sqrt{E_{f}E_{i}}}\right)\left\{\rule{0mm}{7mm}e^{-\beta/2}\frac{d}{d\mu}I\left(\frac{\beta-\alpha}{2\sqrt{\alpha}}\right)-
	e^{\beta/2}\frac{d}{d\mu}I\left(\frac{\beta+\alpha}{2\sqrt{\alpha}}\right)\right\} & = & 
	    \frac{\exp\left\{-(\alpha+\beta^{2}/\alpha)/4\right\}}{\sqrt{\alpha}}
\end{eqnarray}
whereupon
\begin{eqnarray}
\int_{-1}^{\,1}\frac{\exp\left\{-\left(\rule{0mm}{3mm}\alpha+\beta^{2}/\alpha\right)\!/4\right\}}{\sqrt{\alpha}}\,d\mu & = &
\left(\frac{A}{\sqrt{E_{f}E_{i}}}\right)\left[\rule{0mm}{8mm}e^{-\beta/2}\left\{\rule{0mm}{7mm}I\left(\frac{\beta-\alpha_{-}}{2\sqrt{\alpha_{-}}}\right)-
        I\left(\frac{\beta-\alpha_{+}}{2\sqrt{\alpha_{+}}}\right)\right\}-  \right. \nonumber  \\
  &   & \rule{2.4cm}{0mm} \left. e^{\beta/2}\left\{\rule{0mm}{7mm}I\left(\frac{\beta+\alpha_{-}}{2\sqrt{\alpha_{-}}}\right)-
            I\left(\frac{\beta+\alpha_{+}}{2\sqrt{\alpha_{+}}}\right)\right\}\right]   
\end{eqnarray}
\newpage
\mbox{    }
\newline
\newline
\newline
with
\begin{eqnarray}
	\alpha_{\pm} & = & \frac{E_{f}+E_{i}\pm 2\sqrt{E_{f}E_{i}}}{A} \nonumber \\
                 & = & \frac{\left(\sqrt{E_{f}}\pm \sqrt{E_{i}}\,\right)^{2}}{A}
\end{eqnarray}
and
\begin{eqnarray}                                        
  \sqrt{\alpha_{\pm}} & = & \frac{|\sqrt{E_{f}}\pm\sqrt{E_{i}}\,|}{\sqrt{A}} \,,                
\end{eqnarray}
the absolute value bars in the latter being obligatory by virtue of the fact that both orders $E_{i}<E_{f}$ and $E_{i}>E_{f}$
remain in play.  Accordingly, it is only both sign choices of
\[I\left(\frac{\beta\pm \alpha_{-}}{2\sqrt{\alpha_{-}}}\right)\]
which are subject to said order discrimination, whereas their counterparts     
\[I\left(\frac{\beta\pm \alpha_{+}}{2\sqrt{\alpha_{+}}}\right)\]
admit a seamless evaluation.  One finds

\parindent=0in
{\bf{\underline{\mbox{\boldmath$E_{i}<E_{f}$}}:}}
\parindent=0.5in
\begin{eqnarray}
I\left(\frac{\beta + \alpha_{-}}{2\sqrt{\alpha_{-}}}\right) & = &
  I\left(\rule{0mm}{7mm}\frac{\left(\sqrt{E_{f}}+\sqrt{E_{i}}\,\right)\!\left(\sqrt{E_{f}}-\sqrt{E_{i}}\,\right)+\left(\sqrt{E_{f}}-\sqrt{E_{i}}\,\right)^{2}\!\!/A}
                  {2\left(\sqrt{E_{f}}-\sqrt{E_{i}}\,\right)\!/\sqrt{A}}\right)  \nonumber  \\
    & = & I\left(\rule{0mm}{7mm}\left(\frac{A+1}{2\sqrt{A}}\right)\!\sqrt{E_{f}}+\left(\frac{A-1}{2\sqrt{A}}\right)\!\sqrt{E_{i}}\right)             
\end{eqnarray}
\begin{eqnarray}
	I\left(\frac{\beta - \alpha_{-}}{2\sqrt{\alpha_{-}}}\right) & = &
	I\left(\rule{0mm}{7mm}\frac{\left(\sqrt{E_{f}}+\sqrt{E_{i}}\,\right)\!\left(\sqrt{E_{f}}-\sqrt{E_{i}}\,\right)-\left(\sqrt{E_{f}}-\sqrt{E_{i}}\,\right)^{2}\!\!/A}
	{2\left(\sqrt{E_{f}}-\sqrt{E_{i}}\,\right)\!/\sqrt{A}}\right)  \nonumber  \\
	& = & I\left(\rule{0mm}{7mm}\left(\frac{A-1}{2\sqrt{A}}\right)\!\sqrt{E_{f}}+\left(\frac{A+1}{2\sqrt{A}}\right)\!\sqrt{E_{i}}\right)             
\end{eqnarray}
\parindent=0in
{\bf{\underline{\mbox{\boldmath$E_{i}>E_{f}$}}:}}
\parindent=0.5in
\begin{eqnarray}
	I\left(\frac{\beta + \alpha_{-}}{2\sqrt{\alpha_{-}}}\right) & = & \rule{3mm}{0mm}
	I\left(\rule{0mm}{7mm}\frac{-\left(\sqrt{E_{f}}+\sqrt{E_{i}}\,\right)\!\left(\sqrt{E_{i}}-\sqrt{E_{f}}\,\right)+\left(\sqrt{E_{i}}-\sqrt{E_{f}}\,\right)^{2}\!\!/A}
	{2\left(\sqrt{E_{i}}-\sqrt{E_{f}}\,\right)\!/\sqrt{A}}\right)  \nonumber  \\
	& = &-I\left(\rule{0mm}{7mm}\left(\frac{A-1}{2\sqrt{A}}\right)\!\sqrt{E_{i}}+\left(\frac{A+1}{2\sqrt{A}}\right)\!\sqrt{E_{f}}\right)          
\end{eqnarray}
\newpage
\mbox{    }
\newline
\newline
\begin{eqnarray}
	I\left(\frac{\beta - \alpha_{-}}{2\sqrt{\alpha_{-}}}\right) & = &\rule{3mm}{0mm}
	I\left(\rule{0mm}{7mm}\frac{-\left(\sqrt{E_{f}}+\sqrt{E_{i}}\,\right)\!\left(\sqrt{E_{i}}-\sqrt{E_{f}}\,\right)-\left(\sqrt{E_{i}}-\sqrt{E_{f}}\,\right)^{2}\!\!/A}
	{2\left(\sqrt{E_{i}}-\sqrt{E_{f}}\,\right)\!/\sqrt{A}}\right)  \nonumber  \\
	& = & -I\left(\rule{0mm}{7mm}\left(\frac{A+1}{2\sqrt{A}}\right)\!\sqrt{E_{i}}+\left(\frac{A-1}{2\sqrt{A}}\right)\!\sqrt{E_{f}}\right) \,,         
\end{eqnarray}
in both entries (37)-(38) of which we have utilized the antisymmetry of $I(\xi)$ with respect to its argument.  Indiscriminately valid by contrast as to the order of $E_{f}$
versus $E_{i}$ are the entries
     \begin{eqnarray}
     	I\left(\frac{\beta + \alpha_{+}}{2\sqrt{\alpha_{+}}}\right) & = &
     	I\left(\rule{0mm}{7mm}\frac{\left(\sqrt{E_{f}}+\sqrt{E_{i}}\,\right)\!\left(\sqrt{E_{f}}-\sqrt{E_{i}}\,\right)+\left(\sqrt{E_{f}}+\sqrt{E_{i}}\,\right)^{2}\!\!/A}
     	{2\left(\sqrt{E_{f}}+\sqrt{E_{i}}\,\right)\!/\sqrt{A}}\right)  \nonumber  \\
     	& = & I\left(\rule{0mm}{7mm}\left(\frac{A+1}{2\sqrt{A}}\right)\!\sqrt{E_{f}}-\left(\frac{A-1}{2\sqrt{A}}\right)\!\sqrt{E_{i}}\right)             
     \end{eqnarray}
 and
 \begin{eqnarray}
 	I\left(\frac{\beta - \alpha_{+}}{2\sqrt{\alpha_{+}}}\right) & = &
 	I\left(\rule{0mm}{7mm}\frac{\left(\sqrt{E_{f}}+\sqrt{E_{i}}\,\right)\!\left(\sqrt{E_{f}}-\sqrt{E_{i}}\,\right)-\left(\sqrt{E_{f}}+\sqrt{E_{i}}\,\right)^{2}\!\!/A}
 	{2\left(\sqrt{E_{f}}+\sqrt{E_{i}}\,\right)\!/\sqrt{A}}\right)  \nonumber  \\
 	& = & I\left(\rule{0mm}{7mm}\left(\frac{A-1}{2\sqrt{A}}\right)\!\sqrt{E_{f}}-\left(\frac{A+1}{2\sqrt{A}}\right)\!\sqrt{E_{i}}\right) \,.            
 \end{eqnarray}

     Putting it all together, we find that (29) and (32), in conjunction with (35)-(40) and a belated reference perhaps to
     abbreviations (3) give
     
     \parindent=0in
     {\bf{\underline{\mbox{\boldmath$E_{i}<E_{f}$}}:}}
     \parindent=0.5in
     \begin{eqnarray}
     \left\langle
     \frac{d\sigma(\mbox{\boldmath$p$}_{i}\rightarrow \mbox{\boldmath$p$}_{f})}
     {dE_{f}}\right\rangle_{\!\!T} & = & \frac{\sigma_{0}\theta^{2}}{\sqrt{\pi}E_{i}}\left(\rule{0mm}{8mm}
       e^{\left(E_{i}-E_{f}\right)}\left[\rule{0mm}{6mm}
       I\left(\rule{0mm}{5mm}\theta\sqrt{E_{i}}-\zeta\sqrt{E_{f}}\right)+I\left(\rule{0mm}{5mm}\theta\sqrt{E_{i}}+\zeta\sqrt{E_{f}}\right)\right]+ \right.\nonumber  \\
       &   & \rule{2.8cm}{0mm}\left.\rule{0mm}{8mm}\left[\rule{0mm}{6mm}I\left(\rule{0mm}{5mm}\theta\sqrt{E_{f}}-\zeta\sqrt{E_{i}}\right)-
                               I\left(\rule{0mm}{5mm}\theta\sqrt{E_{f}}+\zeta\sqrt{E_{i}}\right)\right]\right)     	
     \end{eqnarray}	
\parindent=0in
{\bf{\underline{\mbox{\boldmath$E_{i}>E_{f}$}}:}}
\parindent=0.5in
\begin{eqnarray}
	\left\langle
	\frac{d\sigma(\mbox{\boldmath$p$}_{i}\rightarrow \mbox{\boldmath$p$}_{f})}
	{dE_{f}}\right\rangle_{\!\!T} & = & \frac{\sigma_{0}\theta^{2}}{\sqrt{\pi}E_{i}}\left(\rule{0mm}{8mm}
	e^{\left(E_{i}-E_{f}\right)}\left[\rule{0mm}{6mm}
	I\left(\rule{0mm}{5mm}\theta\sqrt{E_{i}}-\zeta\sqrt{E_{f}}\right)-I\left(\rule{0mm}{5mm}\theta\sqrt{E_{i}}+\zeta\sqrt{E_{f}}\right)\right]+ \right.\nonumber  \\
	&   & \rule{2.8cm}{0mm}\left.\rule{0mm}{8mm}\left[\rule{0mm}{6mm}I\left(\rule{0mm}{5mm}\theta\sqrt{E_{f}}-\zeta\sqrt{E_{i}}\right)+
	I\left(\rule{0mm}{5mm}\theta\sqrt{E_{f}}+\zeta\sqrt{E_{i}}\right)\right]\right)\,.     	
\end{eqnarray}
\newpage
\mbox{   }
\newline
\newline
\newline
Save for some obvious changes in notation, a passage from unnormalized to normalized error functions, and a promotion of the
microscopic cross section $\sigma_{0}$ (having a dimension of $distance^{2}$) to its macroscopic counterpart $\Sigma_{f}$
(with $distance^{-1}$ as its dimension), obtained by multiplication of $\sigma_{0}$ by the background thermalizing density,
Eqs. (41)-(42) are in complete accord with the composite Eq. (2.19a) as found on p. 26 of [5].
\section{Reconciliation with the classical W-W scattering kernel (2)}
    Under a classical perspective one of course sets $E=mv^{2}/2kT.$  Furthermore, a shift of emphasis from energy to
    the accompanying velocity prompts us to replace the derivatives with respect to energy $d/dE_{f}$ on the left in
    (41)-(42) by $(1/mv_{f})d/dE_{f}.$  And lastly, use of the macroscopic cross section in Boltzmann's neutron
    transport equation (2.1) on p. 14 in [5] requires a preliminary multiplication by $v_{i}.$  Once all reference to
    cross sections, microscopic or otherwise, has been duly removed from (41)-(42) so amended, we are left with the
    intended counterpart to our (2).  Thus
    
    \parindent=0in
    {\bf{\underline{\mbox{\boldmath$v_{i}<v_{f}$}}:}}
    \parindent=0.5in
    \begin{eqnarray}
    	{\underbar{P}}(v_{f},v_{i}) & = & \frac{2\theta^{2}v_{f}}{\sqrt{\pi}v_{i}}\left(\rule{0mm}{8mm}
    	e^{m\left(v_{i}^{2}-v_{f}^{2}\right)/2kT}\left[\rule{0mm}{6mm}
    	I\left(\rule{0mm}{5mm}\frac{\theta v_{i}-\zeta v_{f}}{\sqrt{2kT/m}}\right)+
    	I\left(\rule{0mm}{5mm}\frac{\theta v_{i}+\zeta v_{f}}{\sqrt{2kT/m}}\right)\right]+ \right.\nonumber  \\
    	&  & \rule{3.65cm}{0mm}\rule{0mm}{8mm}\left.\left[\rule{0mm}{6mm}I\left(\rule{0mm}{5mm}\frac{\theta v_{f}-\zeta v_{i}}{\sqrt{2kT/m}}\right)-
    	I\left(\rule{0mm}{5mm}\frac{\theta v_{f}+\zeta v_{i}}{\sqrt{2kT/m}}\right)\right]\right)     	
    \end{eqnarray}	 
   \parindent=0in
   {\bf{\underline{\mbox{\boldmath$v_{i}>v_{f}$}}:}}
   \parindent=0.5in
   \begin{eqnarray}
   	 {\underbar{P}}(v_{f},v_{i}) & = & \frac{2\theta^{2}v_{f}}{\sqrt{\pi}v_{i}}\left(\rule{0mm}{8mm}
   	e^{m\left(v_{i}^{2}-v_{f}^{2}\right)/2kT}\left[\rule{0mm}{6mm}
   	I\left(\rule{0mm}{5mm}\frac{\theta v_{i}-\zeta v_{f}}{\sqrt{2kT/m}}\right)-
   		I\left(\rule{0mm}{5mm}\frac{\theta v_{i}+\zeta v_{f}}{\sqrt{2kT/m}}\right)\right]+ \right.\nonumber  \\
   	&   & \rule{3.65cm}{0mm}\left.\rule{0mm}{8mm}\left[\rule{0mm}{6mm}I\left(\rule{0mm}{5mm}\frac{\theta v_{f}-\zeta v_{i}}{\sqrt{2kT/m}}\right)+
   	I\left(\rule{0mm}{5mm}\frac{\theta v_{f}+\zeta v_{i}}{\sqrt{2kT/m}}\right)\right]\right)\,,     	
   \end{eqnarray}
which, happily enough, is (2) once more.  And so we are done.  One notes in passing that in (2), (41)-(42), and (43)-(44), it is only the interstitial
signs which need to be toggled when one passes from an upscattering to a downscattering kernel.
\section{A highly opinionated coda on dimensions}
    One may observe that we have been at pains to place into evidence fully dimensionless quantities through an overt division
    by the appropriate parameter clusters, for instance, velocities divided by \mbox{$\sqrt{2kT/m}$ in (43)}
    \newpage
    \mbox{   }
    \newline
    \newline
    \newline
    and (44).  The corresponding
    attitude of the authors in [1] and [5] has, by contrast, hewed to the standard, if unnervingly cavalier route prevalent in scientific discourse, of
    simply legislating that certain parameters are to be regarded as one (1!), and never mind the fate of the units involved, with all physical consequences
    to be sorted out somehow only at calculation's end.  For instance, in [1], the authors set the neutron mass $m=1,$ whereas in [5], author
    M. M. R. Williams throws all caution to the winds and sets $\hbar=k=n=1$ {\emph{en masse}} (footnote, p. 17).  No one, evidently, is going to argue that
    diminishing symbol clutter is not all to the good.  But it would seem to me that it should be done honestly and in a controlled fashion, by
    grouping the available parameters of the mathematical context at hand, singly or in suitable clusters, and then rendering the dynamic sinews of the
    physical theory dimensionless through mere division.  The dynamical equations, when finally solved in this dimensionless, universal setting,
    provide, through said parameter groupings, a vista upon a continuum of kindred physical scenarios.  Essentially arbitrary control over the
    relevant parameters can indeed be likened to the power of a puppeteer.
    
        Opportunities for parametric clustering abound in physics.  As one elementary example we may note that Dirac's equation
        readily admits having space measured off in units of the reduced Compton wavelength $\hbar/m_{e}c,$ with
        $m_{e}$ being the electron mass.  In essentially the same breath its time can be gauged in units of $\hbar/m_{e}c^{2}.$
        By way of a somewhat more prosaic example, time-harmonic Maxwell fields $\mbox{\boldmath$E$}$ and
        $\mbox{\boldmath$B$}$ are universally viewed when space is measured in units of wavelength $\lambda=2\pi c/\omega,$ with $\omega$ being the
        angular frequency.  And so on.
        
        In elementary particle physics especially one finds preposterous, outrageous statements such as {\emph{``energy = frequency = mass, and all
        become inverse lengths''}} [11], and reaffirmed in [12], uttered by authors of highly respectable pedigrees.  One is forced then to endure
    the queasy feeling of being entangled by quantities of suspect dimension.  Far better to cluster prudently and render dimensionless through division
    in advance.      

\section{Comments}
    The notes on which this essay is based were assembled many a moon ago, when I was already
    neither young but not yet old.  Buoyed by the gusto of the time, it had seemed to me then that
    the quantum-mechanical route to the W-W scattering kernel was the easier of the two, superior to its
    classical standby.  But now I am not so sure.  Certainly the present material is a bewildering
    mix of the sacred and the profane as regards its attitude toward mathematical rectitude, let
    alone rigor.
    
    In any event, the notes ended just at the point of the final gasp, the integration (29) over the
    angle, or, more precisely, the cosine $\mu$ thereof, separating initial and
    final neutron velocities/momenta.  I had allowed myself to be lulled into thinking all along that
    this integration would be routine and, upon a chance discovery that it was not really so, I was led to
    re{\"{e}}xamine the entire kernel matter, and was thus irresistibly drawn into the chore of
    composition, to crossing every $t$ and dotting every $i.$
    
    All in all I am very glad that I had retained these notes.  I seriously doubt that, at this
    point in time, I would have the tenacity to regenerate them {\emph{de novo.}}
\newpage
\mbox{  }
\section{References}
\vspace{-3mm}
\parindent=0.0in
1.	E. P. Wigner and J. E. Wilkins, Jr., {\bf{Effect of the temperature of the moderator on the velocity distribution of neutrons
with numerical calculations for H as moderator}}, Atomic Energy Commission, Division of Technical Information, Oak Ridge,
Tennessee, Document 2275 (1944).

2.	Kaye Don Lathrop, {\bf{Neutron thermalization in solids}}, California Institute of Technology, Pasadena, PhD Dissertation,
Appendix A (1962).

3.	Gerald C. Pomraning, unpublished handwritten notes entrusted to the undersigned by their late author, copies available upon request.

4.	J. Grzesik, {\bf{Wigner-Wilkins thermalization revisited}}, Nuclear Science and Engineering, Technical Note, Volume 68,
pp. 91-94 (1978).

5.	M. M. R. Williams, {\bf{The slowing down and thermalization of neturons}}, American Elsevier Publishing Company, Inc.,
New York, pp. 26-28 (1966).

6.	John M. Blatt and Victor F. Weisskopf, {\bf{Theoretical nuclear physics}}, Springer-Verlag, New York, pp. 56-86 (1979).

7.	L. D. Landau and E. M. Lifshitz, {\bf{Quantum Mechanics}}, Volume 3 of {\em{Course of Theoretical Physics}}, Third
Edition, Pergamon Press, New York, pp. 640-644 (1977).

8.	G. L. Squires, {\bf{Introduction to the theory of thermal neutron scattering}}, Dover Publications, Inc., Mineola,
New York, pp. 15-18 (1978).

9.	James D. Bjorken and Sidney D. Drell, {\bf{Relativistic quantum mechanics}}, McGraw-Hill Book Company, New York, p. 101
(1964).

10.	Francis E. Low (notes by Wayne A. Mills), {\bf{Quantum Theory of Scattering}}, {\emph{ Lecture Notes:  Summer Institute
in Theoretical Physics}}, Brandeis University, Waltham, Massachusetts, pp. 5-79 (especially 5-12), 1959.

11.	Walter E. Thirring, {\bf{Principles of quantum electrodynamics}}, Academic Press, New York, p. 3 (1958).

12.	T. D. Lee, {\bf{Particle physics and introduction to field theory}}, Harwood Academic Publishers, New York, pp. 1-2 (1981).

\parindent=0.0in
References [1]-[2] are easily retrieved, {\emph{gratis,}} from the internet.


\end{document}